\documentclass{emulateapj}
\usepackage{graphics,graphicx}
\usepackage{helvet}
\usepackage[utf8]{inputenc}
\usepackage[T1]{fontenc}
\usepackage{longtable}
\usepackage{soul}
\usepackage{hyperref}
\usepackage{amsmath, amssymb, gensymb}
\usepackage{ulem}
\usepackage{natbib}
\usepackage{microtype}
\usepackage{multirow}
\newcommand{\kms}{km~s$^{-1}$}

\newcommand{\etal}{{\it et al.\ }}

\newcommand{\srcobs}{27}
\newcommand{\srcnum}{16}
\newcommand{\dbloutflow}{\rm Two}
\newcommand{\bfieldoutflow}{19}

\newcommand{\kstwenty}{$3 \times 10^{-9}$}
\newcommand{\ksrand}{0.64}
\newcommand{\ksanti}{0.79}

\begin{document}

\slugcomment{{\it Published in ApJ on 2013 April 25: ApJ, 786, 159.}}

\title{Misalignment of Magnetic Fields and Outflows in Protostellar Cores}
\shorttitle{Misalignment of Magnetic Fields and Outflows in Protostellar Cores}

\author{Charles L. H. Hull\altaffilmark{1}, Richard L. Plambeck\altaffilmark{1}, Alberto D. Bolatto\altaffilmark{5}, Geoffrey C. Bower\altaffilmark{1},
John M. Carpenter\altaffilmark{2}, Richard M. Crutcher\altaffilmark{4}, Jason D. Fiege\altaffilmark{9}, Erica Franzmann\altaffilmark{9}, 
Nicholas S. Hakobian\altaffilmark{4}, Carl Heiles\altaffilmark{1}, Martin Houde\altaffilmark{10,3},
A. Meredith Hughes\altaffilmark{1,14}, Katherine Jameson\altaffilmark{5}, Woojin Kwon\altaffilmark{4,13}, James W. Lamb\altaffilmark{2}, 
Leslie W. Looney\altaffilmark{4,7}, Brenda C. Matthews\altaffilmark{11,12}, Lee Mundy\altaffilmark{5},
Thushara Pillai\altaffilmark{2}, Marc W. Pound\altaffilmark{5}, Ian W. Stephens\altaffilmark{4}, John J. Tobin\altaffilmark{7,15},
John E. Vaillancourt\altaffilmark{8}, N. H. Volgenau\altaffilmark{6}, and Melvyn C. H. Wright\altaffilmark{1}}

\shortauthors{Hull \etal}
\email{chat@astro.berkeley.edu}

\altaffiltext{1}{Astronomy Department \& Radio Astronomy Laboratory, University of California, Berkeley, CA 94720-3411}
\altaffiltext{2}{Department of Astronomy, California Institute of Technology, 1200 E. California Blvd., MC 249-17, Pasadena, CA 91125, USA}
\altaffiltext{3}{Division of Physics, Mathematics, \& Astronomy, California Institute of Technology, Pasadena, CA 91125, USA}
\altaffiltext{4}{Department of Astronomy, University of Illinois at Urbana-Champaign, 1002 W Green Street, Urbana, IL 61801, USA}
\altaffiltext{5}{Astronomy Department \& Laboratory for Millimeter-wave Astronomy, University of Maryland, College Park, MD 20742}
\altaffiltext{6}{Combined Array for Research in Millimeter-wave Astronomy, Owens Valley Radio Observatory, P.O. Box 968, Big Pine, CA 93513, USA}
\altaffiltext{7}{National Radio Astronomy Observatory, 520 Edgemont Rd., Charlottesville, VA 22903, USA}
\altaffiltext{8}{SOFIA Science Center, Universities Space Research Association, NASA Ames Research Center, Moffett Field, CA 94035, USA}
\altaffiltext{9}{Department of Physics \& Astronomy, University of Manitoba, Winnipeg, MB, R3T 2N2, Canada}
\altaffiltext{10}{Department of Physics \& Astronomy, University of Western Ontario, London, ON, N6A 3K7, Canada}
\altaffiltext{11}{Department of Physics \& Astronomy, University of Victoria, 3800 Finnerty Rd., Victoria, BC, V8P 5C2, Canada}
\altaffiltext{12}{National Research Council of Canada, 5071 West Saanich Rd., Victoria, BC, V9E 2E7, Canada}
\altaffiltext{13}{SRON Netherlands Institute for Space Research, Landleven 12, 9747 AD Groningen, The Netherlands}
\altaffiltext{14}{Miller Fellow}
\altaffiltext{15}{Hubble Fellow}

\begin{abstract}
\noindent
We present results of $\lambda1.3$ mm dust polarization observations toward 
$\srcnum$ nearby, low-mass protostars, mapped with $\sim$2.5$\arcsec$ resolution at CARMA. 
The results show that magnetic fields in protostellar 
cores on scales of $\sim$1000~AU are not tightly aligned with outflows from the protostars. 
Rather, the data are consistent with scenarios where outflows and magnetic fields
are preferentially misaligned (perpendicular), or where they are randomly aligned.
If one assumes that outflows emerge along the rotation axes of circumstellar disks, 
and that the outflows have not disrupted the fields in the surrounding material,
then our results imply that the disks are not aligned with the fields in the cores from which they formed.





%
\end{abstract}

\keywords{ISM: magnetic fields --- magnetic fields --- polarization --- stars: formation --- stars: magnetic field --- stars: protostars}

\section{INTRODUCTION}
\label{sec:intro}

Optical polarization measurements of background stars show that magnetic fields
are well ordered on large scales in the low-density envelopes of molecular
clouds, as in the Pipe Nebula \citep{Franco2010}, which suggests that these
parsec-scale envelopes are magnetically supported (``subcritical'').
Ultimately, however, ambipolar diffusion \citep[\textit{e.g.,}][]{Fiedler1993}, 
turbulence \citep[\textit{e.g.,}][]{MacLow2004}, or turbulent magnetic
reconnection diffusion \citep{Lazarian2005, Leao2012} allows the formation of
``supercritical'' dense cores in which gravity overwhelms magnetic support.  In
the simplest axisymmetric case, one expects the field lines to be drawn
into an hourglass shape by gravitational collapse, forming a $\sim$1000~AU diameter
``pseudodisk'' \citep{Galli1993}.

Although the magnetic field (B-field) may not be strong enough to prevent the formation
of a protostar within the pseudodisk, it can have a significant impact on the accretion
rate onto the star, and on the formation of a rotationally supported
circumstellar disk in which planets will form.  In the limiting case of flux
freezing, the field close to the protostar becomes strong enough to brake the
rotation of the infalling gas completely, preventing the formation of a
rotationally supported disk \citep{Galli2006}.  Even if ambipolar diffusion
allows the field to escape the central core, the redistributed flux tends to be
trapped in a ring surrounding the star, greatly reducing the infall rate
\citep{Li2011}.  

Magnetic braking is less effective, and disks should form more easily, if the
rotation axis of the cloud is at an angle to the magnetic field
\citep{Joos2012, Krumholz2013}.  Indeed, from an observational perspective it is clear that
circumstellar disks typically are {\it not} well-aligned with
the parsec-scale magnetic fields in the surrounding molecular cloud.  If they were, the
bipolar outflows and jets that emerge along the axes of these disks would all
be parallel with the ambient B-field, but this is not observed: \citet{Menard2004} have shown that the
optical jets from classical T Tauri stars in the Taurus-Auriga molecular cloud
are randomly oriented with respect to the parsec-scale magnetic field in this
cloud; \citet{Targon2011} obtained a similar result for 28 regions
spread over the Galaxy, although they do find some evidence for alignment of
jets from younger, Class 0 and Class I, protostars.

Polarization observations of background stars are unable to probe the magnetic
field morphologies inside the dense cores where circumstellar disks form; even at
infrared wavelengths the extinction through these regions is too great.
Mapping the polarized thermal emission from dust grains at millimeter and submillimeter
wavelengths is the usual means of studying the magnetic fields in these
regions.  Under most circumstances, spinning dust grains are expected to align
themselves with their long axes perpendicular to the magnetic field
\citep{Hoang2009}, so normally the thermal radiation from these grains is
weakly polarized perpendicular to the field.

Dust polarization maps of many sources have been obtained at submillimeter
wavelengths with single-dish instruments, {\it e.g.,} with the SCUBA polarimeter 
on the JCMT at 850 $\mu$m \citep{Matthews2009}, and with the Hertz polarimeter 
on the CSO at 350 $\mu$m \citep{Dotson2010}.  These maps have
angular resolutions of about $20''$, corresponding to scales of 3000--8000~AU
in nearby molecular clouds.  \citet{Curran2007} found that on these scales, outflows
and inferred B-fields are randomly aligned.


Higher angular resolution is required to study the field geometry in the
innermost regions of the cores where circumstellar disks form.  Thus far,
interferometric polarization maps with resolutions of a few arcseconds 
have been published for about a dozen sources, using data from OVRO, BIMA, and the SMA.
Most of this work has focused on detailed analyses of individual objects: 
examples include maps of NGC~1333-IRAS~4A \citep{Girart2006}, IRAS~16293
\citep{Rao2009}, and Orion~KL \citep{Rao1998,Tang2010}.  In this limited set of
sources, outflows often were found to be skewed with respect to the inferred
magnetic field directions, hinting that circumstellar disks may not be tightly
aligned with the magnetic fields on $\sim$1000~AU scales.

The 1~mm dual-polarization receiver system at CARMA (the Combined Array
for Research in Millimeter-wave Astronomy) allows us 
to map the dust polarization toward many more sources.  
Here we present results from the 
TADPOL\footnote{ {\bf T}elescope {\bf A}rray {\bf D}oing {\bf POL}arization: \\ \url{tadpol.astro.illinois.edu/}} 
survey, a CARMA key project to study dust polarization in protostellar cores.  
This paper focuses on results from nearby, low-mass protostars.
We compare the field direction inferred from dust-polarization measurements 
with the outflow direction, which indicates the axis of the rotationally 
supported disk.  On the $\sim$1000~AU scales probed by our data,
magnetic fields appear to be either preferentially misaligned (perpendicular) 
or randomly aligned with respect to outflows. 



\section{OBSERVATIONS AND DATA REDUCTION}
\label{sec:obs_redux}


Observations were made with CARMA between May 2011 and October 2012. 
We selected sources from catalogs of young stellar objects, including
\citet{Jorgensen2007}, \citet{Matthews2009}, \citet{2010ApJ...712.1010T},
and \citet{Enoch2011}.  We focus on Class 0 and Class I objects 
at distances of $d \lesssim 400~{\rm pc}$ that are known to have bipolar outflows.

The polarization system 
consists of dual-polarization receivers that are sensitive to right- (R) and left-circular (L) polarization,
and a spectral-line correlator that measures all four cross polarizations 
(RR, LL, LR, RL) on the 105 baselines connecting the 15 antennas.  
The receivers comprise a single feed horn, a waveguide circular polarizer \citep{Plambeck_2010}, 
an orthomode transducer \citep{TMTT.2005.860505}, and two mixers.  
The receivers are double-sideband; a phase-switching
pattern applied to the local oscillator (LO) allows signals 
in the lower (LSB) and upper sidebands (USB) to be separated in the correlator.



For these observations, the correlator was set up with 
three 500 MHz-wide bands to measure the dust continuum,
and one 31 MHz-wide band to map bipolar outflows.  
The frequency of the first LO was 223.821 GHz.  
The three continuum bands 
were centered at 6.0, 7.5, and 8.0 GHz in the intermediate frequency (IF). 
The corresponding RF frequencies are equal to the difference (LSB) or the 
sum (USB) of the LO and the IF.   
The spectral-line band was centered at IF = 6.717 GHz, allowing simultaneous 
observations of the SiO(5--4) line in the LSB, and the 
 CO(2--1) line in the USB.  For the spectral line measurements, the channel spacing 
is 0.2 \kms.


In addition to the usual gain, passband, 
and flux calibrations, polarization observations 
require two additional calibrations: ``XYphase'' and leakage. 
The XYphase calibration corrects for phase differences between the L and R
receivers, and is done by observing an artificially polarized noise
source of known position angle. 
The leakage corrections compensate for cross-coupling between the L and R receivers, and  
are calibrated by observing a strong source (usually the gain calibrator) over
a range of parallactic angles.  
There are no moving parts in the CARMA dual
polarization receivers, so the measured leakages are stable with time.
A typical antenna has a band-averaged leakage amplitude ({\it i.e.,}
a voltage coupling from L into R, or vice versa) of 6\%.





We perform calibration and imaging with the MIRIAD data reduction package
\citep{1995adass...4..433S}.  
Using multi-frequency synthesis and natural weighting, we create dust-continuum maps of all 
four Stokes parameters $(I, Q, U, V)$.
The typical beam size is 2.5\arcsec, which corresponds to 
a resolution of 750~AU at a distance of 300 pc.
We produce polarization position-angle and intensity maps from the Stokes $I$, $Q$, and $U$ data, 
where the position angle of the incoming radiation is 
$\chi = 0.5 \arctan{\left(U / Q \right)}$, and the bias-corrected polarized intensity is
$P_{\rm c} = \sqrt{Q^2 + U^2 - \sigma_{\rm P}^2}$  \citep{2006PASP..118.1340V}
($\sigma_{\rm P}$ is the median rms noise in the Stokes $Q$ and $U$ maps).

In good weather $\sigma_{\rm P} \approx 0.4$ mJy/beam for a single 6-hour observation, 
and can be as low as $\sim$0.2 mJy/beam when multiple
observations are combined.  We consider it a detection only if 
$P_{\rm c} \geq 3\sigma_{\rm P}$
and the location of the polarized emission coincides with a detection of 
total intensity $I \geq 2\sigma_{\rm I}$, where $\sigma_{\rm I}$ is the rms noise in the 
Stokes $I$ map.

We also generate maps of the red- and blueshifted CO and SiO line wings to measure outflow directions.
We generally use CO maps to measure the outflow direction, but occasionally we use SiO 
if the CO emission is too complex.  We do not attempt to measure polarization in the spectral
line data because of fine-scale frequency structure in the polarization leakages.

\section{RESULTS \& ANALYSIS}
\label{sec:results}

Of the $\srcobs$ TADPOL sources within $\sim$400 pc, we detected
dust polarization toward $\srcnum$, which we
focus on in this paper.  The full set of TADPOL results will be 
presented in a separate paper.  

\begin{figure*} 
\begin{center}
\epsscale{0.38}
\plotone{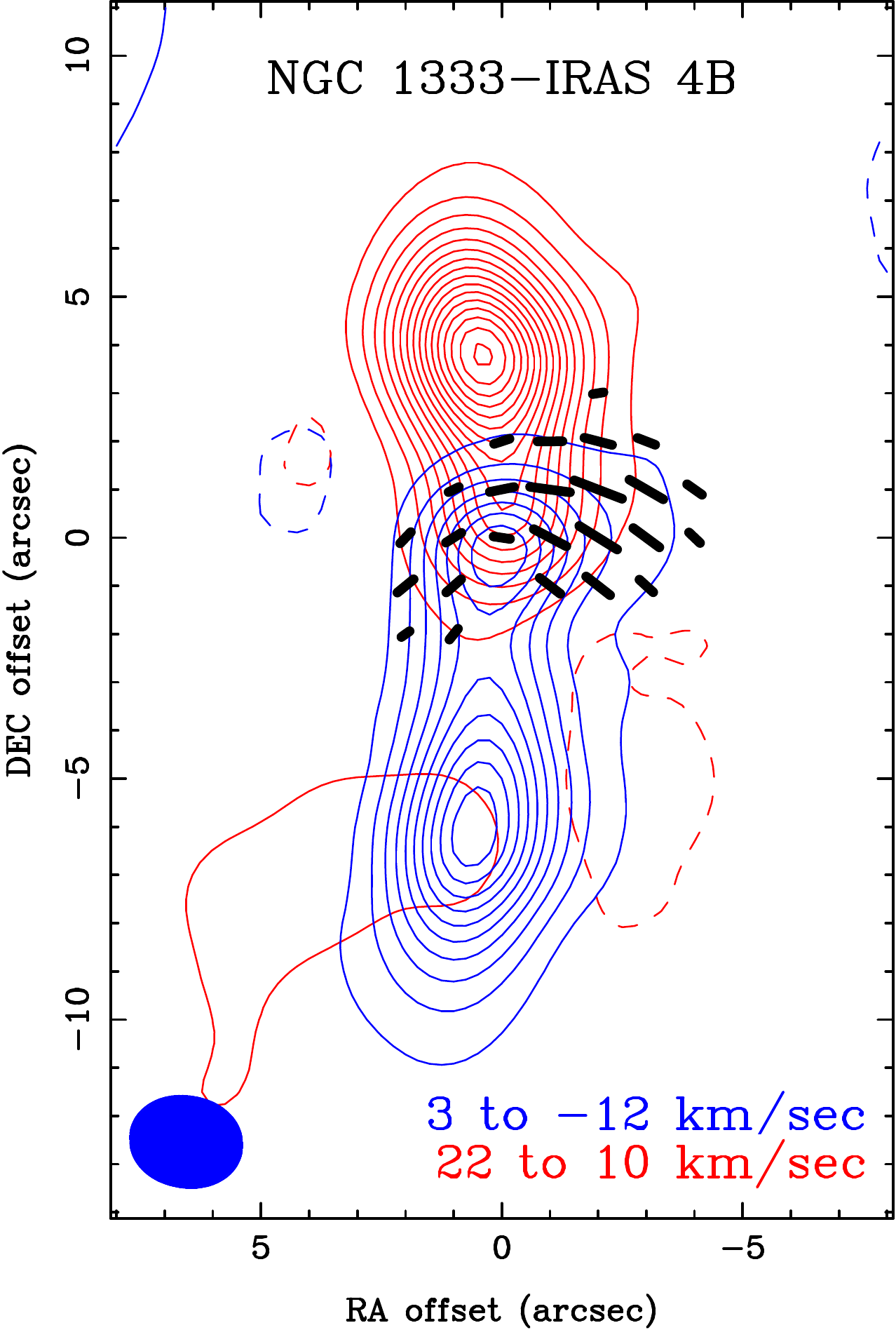}
\epsscale{0.45}
\plotone{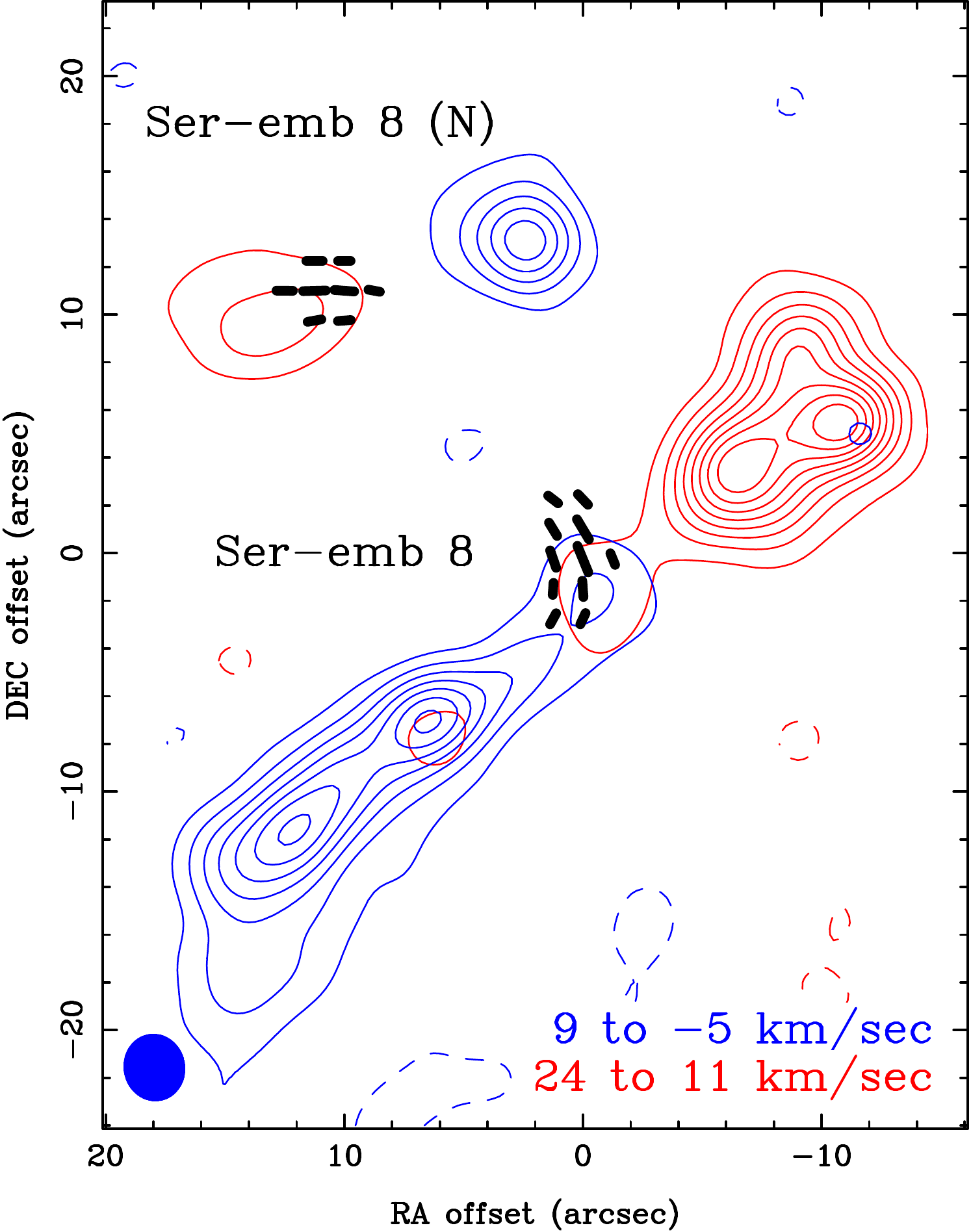}
\epsscale{0.4}
\plotone{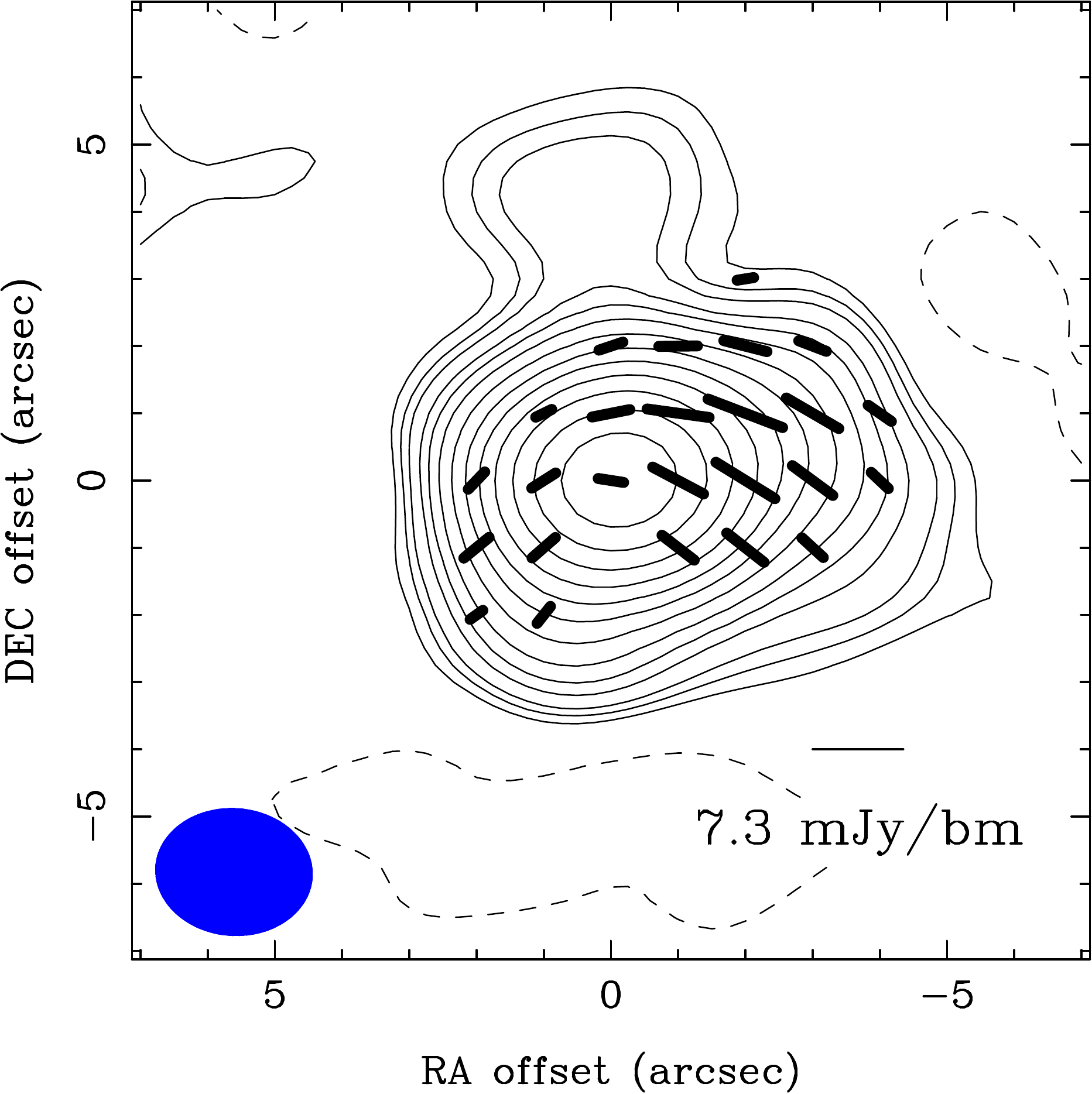}
\epsscale{0.37}
\plotone{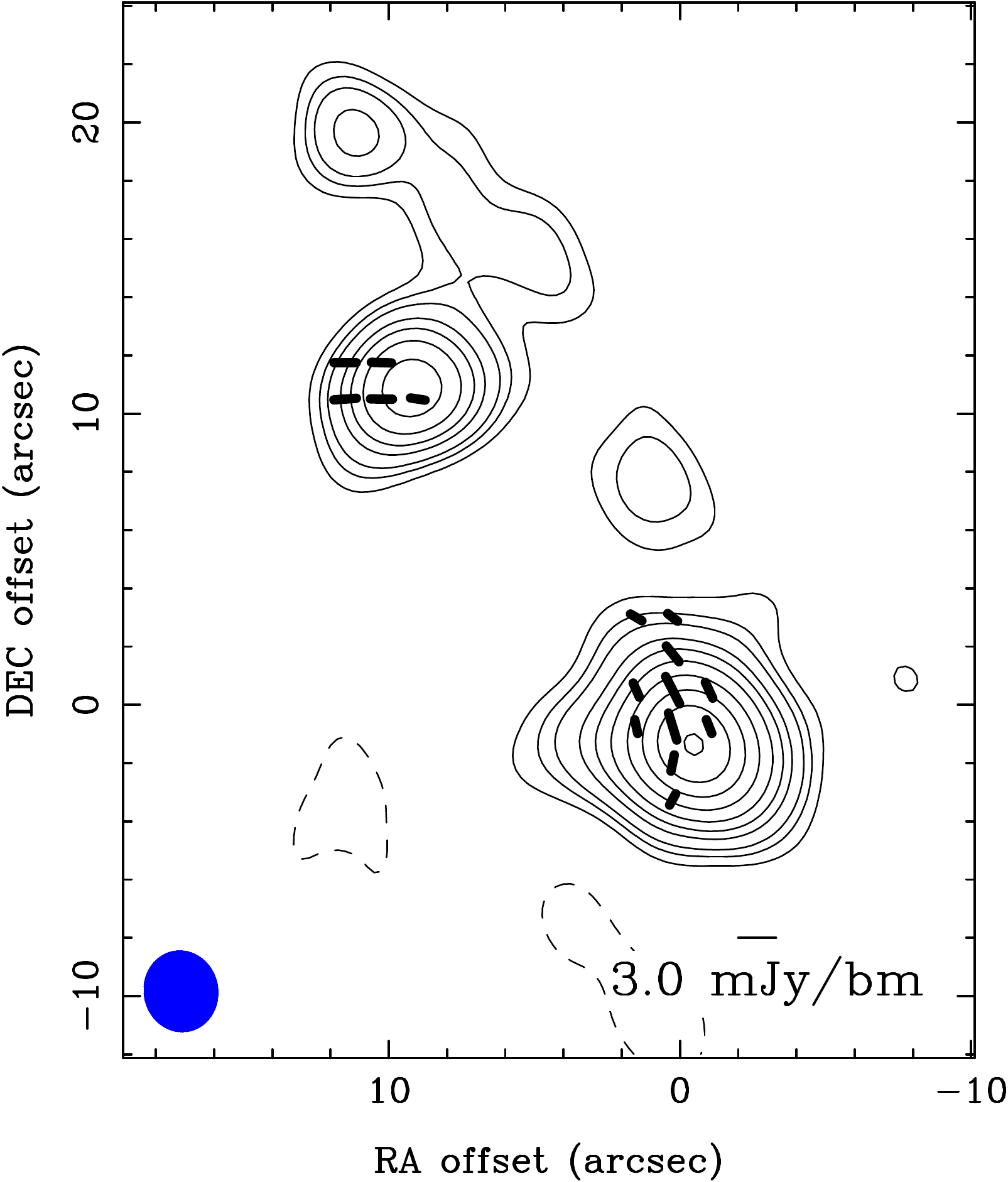}
\caption{ \footnotesize{ 
Sample maps from two sources in the TADPOL survey.  
Line segments show the inferred magnetic field directions; they have been rotated
by $90\degree$ relative to the polarization directions.  
Vector lengths are proportional to polarized intensity, {\it not} fractional 
polarization.   The scale bars show the peak intensities of polarized emission in mJy bm$^{-1}$.
The solid ellipses show the synthesized beams.
{\it Top:} Red- and blue-shifted line wings in CO(2--1) for IRAS~4B, and in SiO(5--4) for Ser-emb~8, 
show the bipolar outflows. Velocity ranges are given in the figures.
Contour levels are in steps of $6\sigma$, beginning at $\pm 4\sigma_{\rm o}$, where $\sigma_{\rm o}$ is the 
rms noise measured in each outflow map.
{\it Bottom:} Dust continuum (Stokes I) contours and inferred B-field
vectors toward the same two sources.  Contours are $-2, 2, 3, 4, 6, 8, 11, 16, 23, 32, 45, 64, 91, 128 \times\sigma_{\rm I}$, 
where $\sigma_{\rm I}$ is the rms noise measured in the dust continuum maps. $\sigma_{\rm I} = 4.3 {\rm ~mJy~bm^{-1}}$
for IRAS~4B; $\sigma_{\rm I} = 3.3 {\rm ~mJy~bm^{-1}}$ for Ser-emb~8. 
}}
\label{fig:sources}
\end{center}
\end{figure*}

Figure \ref{fig:sources} shows some example results.  
In all of the plots, the dust-polarization vectors have been rotated 
by $90\degree$ to show the inferred magnetic field direction and are
plotted at the Nyquist spatial frequency (two vectors per synthesized beam).

 

Table \ref{table:obs} lists the results for the $\srcnum$ TADPOL sources, 
as well as for IRAS~16293, which was 
previously published by \citet{Rao2009}.  $\dbloutflow$ of the sources
each have two distinct outflows, which we consider as independent data, thus making a
total of $\bfieldoutflow$ entries.  Note that NGC~1333-IRAS~4A, one of the sources in 
the TADPOL sample, has been mapped in detail before \citep{Girart2006}, and was 
included in our survey as a cross-check.  The polarization directions 
at the intensity peaks of the 230~GHz CARMA map and the previously published 
345~GHz SMA map are in excellent agreement.

To estimate the outflow direction, we measure the position angles of lines connecting the 
center of the continuum source and the intensity peaks of the red and blue outflow lobes; we take the 
average of the two position angles as the outflow direction.  As a crude estimate of the  
uncertainty $\sigma_{\rm o}$, we use half the difference of the position angles.  
The B-field direction is calculated by averaging the B-field position angles,
weighted by the Stokes I total intensity, which gives the field direction in the densest part of the core.  
The uncertainty in the B-field $\sigma_{\rm B}$ is the average of the B-field position angle uncertainties,
also weighted by the Stokes I total intensity.
The total uncertainty in the angle between the B-field and the outflow is 
$\sigma_{\rm o-B} = \sqrt{ \sigma_{\rm o}^2 + \sigma_{\rm B}^2 }$. 

\begin{figure*}
\begin{center}
\epsscale{.89}
\plotone{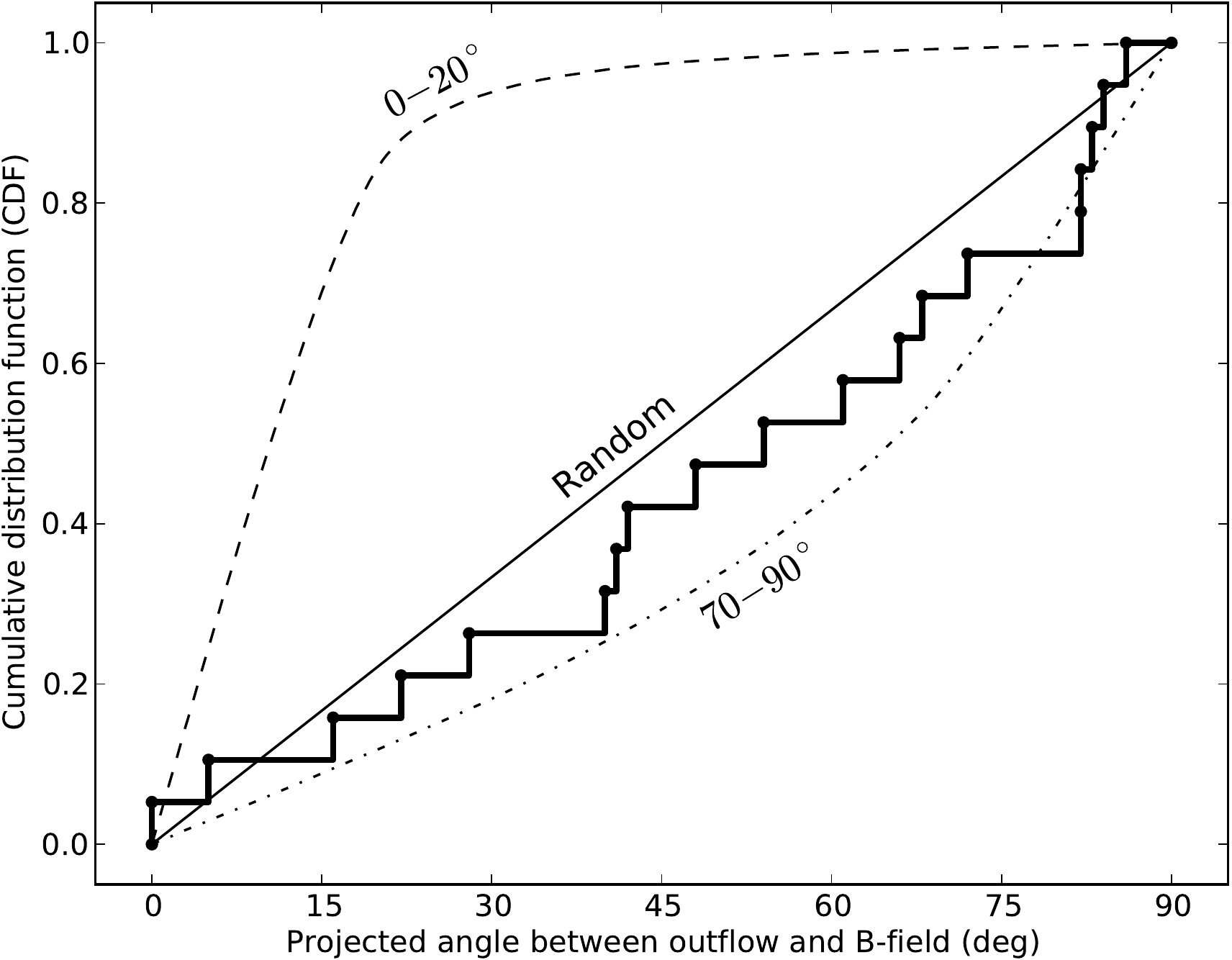}
\caption{The thick solid curve shows the cumulative distribution function (CDF) of the (projected) angles between the mean magnetic field and
outflow directions for the sources in Table \ref{table:obs}. The upper dashed curve is the CDF from a Monte-Carlo simulation 
where outflow and B-field directions are oriented within $20\degree$ of one another (tightly aligned). 
The lower dot-dashed curve is the CDF from a simulation 
where outflow and B-field directions are separated by 70--90$\degree$ (preferentially misaligned).
The straight line is the CDF for random orientation.}
\label{fig:ks}
\end{center}
\end{figure*}

Figure \ref{fig:ks} shows the cumulative distribution function (CDF) of 
the projected angles between the B-fields and outflows of the sources in Table \ref{table:obs}.   
The B-field and outflow position angles we observe are projected onto the plane of the sky.
To determine if the large scatter in position angle differences could be due to projection effects, 
we compare the results with Monte Carlo simulations where the outflows and B-fields are 
tightly aligned, preferentially misaligned (perpendicular), or randomly aligned. 

For the tightly aligned case, the simulation randomly selects pairs of vectors in three dimensions that are within
20$\degree$ of one another, and then projects the vectors onto the plane of the sky
and measures their angular differences.  The resulting CDF is shown by the upper dotted curve in 
Figure \ref{fig:ks}.  In this case projection effects are not as problematic as one might think: to have a
projected separation larger than 20$\degree$, the two vectors must point almost along the line of sight.  

For the preferentially misaligned case, the simulation randomly selects pairs of vectors that are separated
by 70--90$\degree$.  The resulting CDF is shown by the lower dotted curve in 
Figure \ref{fig:ks}.  In this case projection effects are more important, and result in a CDF that is
similar to that expected for random alignment, shown by the solid curve.

A Kolmogorov-Smirnov (K-S) test shows that the probability that our data were drawn from the same population
as the tightly aligned model is \kstwenty, ruling out this scenario.
The probability that the results were drawn from a preferentially misaligned population is \ksanti, 
and from a random population is \ksrand.  Although the probability is slightly higher for the misaligned case,
either of the latter two models are consistent with the data.


\begin{deluxetable*}{lrrrllrcc} 
\tabletypesize{\scriptsize}
\tablecaption{ Observations }
\tablewidth{0pt}
\tablehead{
\colhead{Source} & \colhead{$\alpha$} & \colhead{$\delta$} & \colhead{$\chi_{\rm B}$ $(\sigma_{\rm B})$} &
\colhead{$\chi_{\rm o}$ $(\sigma_{\rm o})$} & \colhead{$\theta_{\rm o-B}$ $(\sigma_{\rm o-B})$} &
\colhead{$\theta_{\rm bm}$} & \colhead{$d$} & \colhead{Distance} \\
                                & \colhead{(J2000)} & \colhead{(J2000)} & \colhead{($\degree$)} & \colhead{($\degree$)} & 
                                \colhead{($\degree$)} & \colhead{($\arcsec$)} & \colhead{(pc)} & \colhead{ref.\tablenotemark{d}}
}

\startdata
L1448~IRS~2 & 03:25:22.4 & 30:45:13.2 & 139 (9) & 134 (5) & \phantom{1}5 (10) & 3.63 & 230 & 1\\
L1448N(B) & 03:25:36.3 & 30:45:14.8 & 31 (6) & \phantom{1}97 (2) & 66 (6) & 2.04 & 230 & 1\\
NGC~1333-IRAS~2A\tablenotemark{a} & 03:28:55.6 & 31:14:37.1 & 82 (6) & \phantom{1}98 (6) & 16 (8) & 3.45 & 230 & 1\\
  &   &   & 82 (6) & \phantom{1}21 (9) & 61 (11) &   &   & 1\\
NGC~1333-IRAS~4A & 03:29:10.5 & 31:13:31.3 & 58 (2) & \phantom{1}18 (10) & 40 (10) & 2.52 & 230 & 1\\
NGC~1333-IRAS~4B & 03:29:12.0 & 31:13:08.1 & 86 (5) & \phantom{11}0 (5) & 86 (7) & 2.09 & 230 & 1\\
HH~211 & 03:43:56.8 & 32:00:50.0 & 164 (6) & 116 (1) & 48 (6) & 3.95 & 230 & 1\\
L1551~IRS~5 & 04:31:44.5 & 18:08:31.5 & 165 (4) & \phantom{1}67 (5) & 82 (6) & 2.18 & 140 & 2\\
L1527 & 04:39:53.9 & 26:03:09.6 & 174 (8) & \phantom{1}92 (7) & 82 (11) & 3.06 & 140 & 2\\
OMC3-MMS5\tablenotemark{b} & 05:35:22.4 & --05:01:14.5 & 52 (9) & \phantom{1}80 (6) & 28 (11) & 3.22 & 415 & 3\\
OMC3-MMS6 & 05:35:23.4 & --05:01:30.6 & 45 (2) & 171 (8) & 54 (8) & 3.22 & 415 & 3\\
VLA~1623 & 16:26:26.4 & --24:24:30.4 & 36 (9) & 120 (5) & 84 (10) & 2.89 & 120 & 4\\
IRAS~16293~A\tablenotemark{a,c} & 16:32:22.9 & --24:28:36.3 & 5 (9) & \phantom{1}77 (9) & 72 (13) & 2.46 & 178 & 5\\
  &   &   & 5 (9) & 143 (9) & 42 (13) &   &   & 5\\
Ser-emb~8 & 18:29:48.1 & 01:16:43.6 & 39 (6) & 107 (1) & 68 (6) & 2.63 & 415 & 6\\
Ser-emb~8~(N) & 18:29:48.7 & 01:16:55.8 & 88 (7) & 129 (2) & 41 (7) & 2.63 & 415 & 6\\
Ser-emb~6 & 18:29:49.8 & 01:15:20.3 & 157 (3) & 135 (3) & 22 (4) & 2.71 & 415 & 6\\
L1157 & 20:39:06.2 & 68:02:16.0 & 146 (4) & 146 (7) & \phantom{1}0 (8) & 2.39 & 250 & 7\\
CB~230 & 21:17:38.7 & 68:17:32.4 & 89 (6) & 172 (4) & 83 (7) & 3.05 & 400 & 8
\enddata

\tablenotetext{a}{Source has two outflows.}
\tablenotetext{b}{Coordinates from \citet{Takahashi2011}}
\tablenotetext{c}{Results from \citet{Rao2009}.}
\tablenotetext{d}{Distance references.  1: \citet{Hirota2011}.  2: \citet{Loinard2007}.   
  3: \citet{Menten2007}.  4: \citet{Loinard2008}.  5: \citet{Imai2007}.  6: \citet{Dzib2010}.  7: \citet{Looney2007}.  8: \citet{Launhardt2010}.}
\tablecomments{Coordinates are fitted positions of dust emission peaks.
The outflow angle $\chi_{\rm o}$ and inferred magnetic-field angle $\chi_{\rm B}$ 
are measured counter-clockwise from north.  
The angle difference $\theta_{\rm o-B}$ between the outflow and the B-field is always
between 0--$90\degree$.  The B-field direction is assumed to be perpendicular to the direction of 
the polarized dust emission.  $d$ is the distance to the source.  $\theta_{\rm bm}$ 
is the geometric mean of the major and minor axes of the synthesized beam.}

\label{table:obs}
\end{deluxetable*}

\section{DISCUSSION}
\label{sec:discussion}

Most analytical models of star formation
assume that the rotation axis of a protostellar core, its magnetic field direction, 
and its outflow direction all are parallel \citep{Shu2000,Konigl2000}.  
Our results appear to contradict this simple picture.


We have assumed that dust grains always are aligned with their long axes perpendicular 
to the magnetic fields, and thus that B-fields are perpendicular to the polarization directions.  
This may not always be the case.
For example, if grains are mechanically aligned by outflows, then the polarization is expected
to be {\it parallel} to the B-field \citep{1952MNRAS.112..215G}.  
Mechanical alignment should affect only a small portion 
of a protostellar core, however: the maximum opening angle of the 
outflows in our sample is $\sim$60$\degree$, corresponding to $\sim$10\% of the 
core volume.  \citet{Lazarian2007} and \citet{Hoang2009} show that grains aligned by radiative torques
can also be stably aligned parallel to the B-field under some conditions, but
this is unlikely in the densest part of a protostellar core.  
Outflows may also affect magnetic field morphologies. 
Because of magnetic tension, the influence of an outflow is not restricted 
to the outflow cavity, and potentially could extend over a significant fraction 
of the core volume.  Additionally, simulations by \citet{Tomisaka2011} have shown that 
B-field morphologies can depend on the outflow launching mechanism:
the B-field is predominantly toroidal in a magnetocentrifugally driven wind, 
and is predominantly poloidal in a jet-driven outflow with entrained molecular material.
These differences in morphology, combined with projection effects, 
could result in random orientations between outflows and B-fields.

Finally, some simulations, such as those by \citet{2009A&A...506L..29H}, \citet{Joos2012},
and \citet{Li2013} suggest that misalignment of the B-field and the core rotation axis
can {\it aid} in the formation of circumstellar disks, given a mass-to-flux ratio
 in the core of $\gtrsim 2$.  \citet{Li2011} find
that disk formation is suppressed in the case where fields and core rotation
axes are parallel, even when non-ideal MHD effects are considered.  Hence, 
these models suggest that misalignment may be a necessary condition for the formation
of disks.



\section{SUMMARY}
\label{sec:summary}

Results from the TADPOL survey show that magnetic fields on scales of 
$\sim$1000~AU are not tightly aligned with protostellar outflows.  
Rather, the data are consistent both with scenarios where outflows and magnetic fields
are preferentially misaligned (perpendicular), and where they are randomly aligned.
If one assumes that outflows emerge along the rotation axes of circumstellar disks, 
and that the outflows have not disrupted the fields in the surrounding material,
then our results imply that the disks are not aligned with the fields in the cores from which they formed.

It could be fruitful to investigate whether alignment correlates with
core rotation, field strength, outflow velocity, multiplicity, or age.
Higher resolution polarization observations with
ALMA will test these correlations at the $\sim$100~AU scale of circumstellar disks.



\acknowledgments

We thank the referee for his/her insightful comments.

C.L.H.H. thanks Chris McKee, Richard Klein, Mark Krumholz, and Andrew Myers
for a helpful discussion, and would like to acknowledge the 
advice and guidance of the members of the Berkeley Radio Astronomy 
Laboratory and the Berkeley Astronomy Department.  

C.L.H.H. acknowledges support from an NSF Graduate 
Fellowship. J.D.F. acknowledges support from an NSERC Discovery grant.
J.J.T. acknowledges support provided by NASA through Hubble Fellowship 
grant \#HST-HF-51300.01-A awarded by the Space Telescope Science Institute, which is 
operated by the Association of Universities for Research in Astronomy, 
Inc., for NASA, under contract NAS 5-26555.  
A.D.B. acknowledges support from a CAREER grant  NSF-AST0955836, NSF-AST1139998,
and a Research Corporation for Science Advancement Cottrell Scholar award.

Support for CARMA construction was derived from the states of California, Illinois, and Maryland, the James 
S. McDonnell Foundation, the Gordon and Betty Moore Foundation, the Kenneth T. and Eileen L. Norris 
Foundation, the University of Chicago, the Associates of the California Institute of Technology, and the National 
Science Foundation. Ongoing CARMA development and operations are supported by the National Science 
Foundation under a cooperative agreement, and by the CARMA partner universities.

\bibliographystyle{apj}

\end{document}